\documentclass[sigconf]{acmart}
\usepackage{adjustbox}
\usepackage{booktabs}  
\usepackage{multirow}  

\AtBeginDocument{%
  }

\begin{document}

\title{GReF: A Unified Generative Framework for Efficient Reranking via Ordered Multi-token Prediction}

\author{Zhijie Lin}
\authornotemark[1]
\affiliation{%
  \institution{Kuaishou Technology}
  \city{Beijing}
  \country{China}}
\email{linzhijie@kuaishou.com}

\author{Zhuofeng Li}
\authornote{Both authors contributed equally to this research.}
\affiliation{%
  \institution{Shanghai University}
  \city{Shanghai}
  \country{China}}
\email{zhuofengli12345@gmail.com}

\author{Chenglei Dai}
\authornote{Corresponding Author}
\affiliation{%
  \institution{Kuaishou Technology}
  \city{Beijing}
  \country{China}}
\email{daichenglei@kuaishou.com}

\author{Wentian Bao}
\affiliation{%
 \institution{Independent}
  \city{Beijing}
  \country{China}}
\email{wb2328@columbia.edu}

\author{Shuai Lin}
\affiliation{%
  \institution{Kuaishou Technology}
  \city{Beijing}
  \country{China}}
\email{shuailin97@gmail.com}

\author{Enyun Yu}
\affiliation{%
 \institution{Independent}
  \city{Beijing}
  \country{China}}
\email{yuenyun@126.com}

\author{Haoxiang Zhang}
\affiliation{%
  \institution{Shanghai University}
  \city{Shanghai}
  \country{China}}
\email{Isaac_GHX@shu.edu.cn}

\author{Liang Zhao}
\affiliation{%
  \institution{Emory University}
  \city{Atlanta, GA, USA}
  \country{USA}}
\email{liang.zhao@emory.edu}

\renewcommand{\shortauthors}{Zhijie Lin et al.}

\begin{abstract}
In a multi-stage recommendation system, reranking plays a crucial role in modeling intra-list correlations among items. A key challenge lies in exploring optimal sequences within the combinatorial space of permutations. Recent research follows a two-stage (generator-evaluator) paradigm, where a generator produces multiple feasible sequences, and an evaluator selects the best one. In practice, the generator is typically implemented as an autoregressive model. However, these two-stage methods face two main challenges. First, the separation of the generator and evaluator hinders end-to-end training. Second, autoregressive generators suffer
from inference efficiency. In this work, we propose a Unified Generative Efficient Reranking Framework (GReF) to address the two primary challenges. Specifically, we introduce Gen-Reranker, an autoregressive generator featuring a bidirectional encoder and a dynamic autoregressive decoder to generate causal reranking sequences. Subsequently, we pre-train Gen-Reranker on the item exposure order for high-quality parameter initialization. To eliminate the need for the evaluator while integrating sequence-level evaluation during training for end-to-end optimization, we propose post-training the model through Rerank-DPO. Moreover, for efficient autoregressive inference, we introduce ordered multi-token prediction (OMTP), which trains Gen-Reranker to simultaneously generate multiple future items while preserving their order, ensuring practical deployment in real-time recommender systems. Extensive offline experiments demonstrate that GReF outperforms state-of-the-art reranking methods while achieving latency that is nearly comparable to non-autoregressive models. Additionally, \textbf{GReF has also been deployed in a real-world video app Kuaishou with over 300 million daily active users, significantly improving online recommendation quality.} 
\end{abstract}



\maketitle
\section{INTRODUCTION}


Multi-stage recommendation systems are widely used on platforms like YouTube and Kuaishou. In the final stage, reranking refines the top candidates by applying specific objectives to further improve recommendation quality. Specifically, it takes the ranking list as input and reorders it by modeling cross-item interactions within the listwise context. Thus, the main challenge in reranking is exploring the optimal sequence within the vast space of permutations.



Extensive research on reranking typically falls into two categories: one-stage and two-stage methods. One-stage methods \cite{ai2018learning,pei2019personalized} take candidate items as input, estimate refined scores for each within the permutation, and rerank them greedily based on these scores. However, one-stage methods encounter an inherent contradiction \cite{feng2021grn,xi2021context}: the reranking operation modifies the permutation, introducing different mutual influences between items compared to the initial arrangement. Consequently, the refined score conditioned on the initial permutation is considered implausible. 

To tackle this challenge, two-stage methods \cite{lin2023discrete, shi2023pier, ren2024non, feng2021revisit} adopt a generator-evaluator framework, where the generator produces multiple candidate sequences and the evaluator selects the optimal one based on an estimated listwise score. Compared to one-stage methods that score items individually, two-stage approaches refine the entire sequence using a sequence-level evaluator, better capturing contextual dependencies and aligning with users’ overall preferences.  

\begin{figure}
    \centering
    \includegraphics[width=\linewidth]{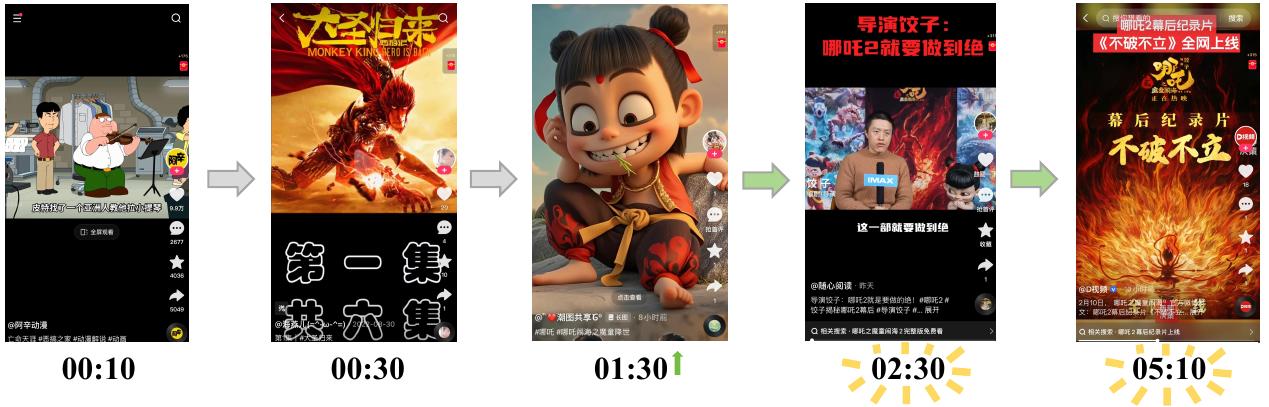}
    \vspace{-5mm}
    \caption{User browsing behavior with causal dependencies}
    \label{fig:compare}
    \vspace{-6mm}
\end{figure}

In practice, the generator is typically implemented as an autoregressive model, such as Seq2Slate \cite{bello2018seq2slate}, to capture the causal dependencies that are widely present in user browsing behavior. As illustrated in Figure \ref{fig:compare}, a user who watches \textit{Monkey King: Hero is Back} may subsequently browse a related Chinese animated film like \textit{Nezha}, followed by additional content from the Nezha series, eventually spending several minutes on the final video. Such widely observed causal browsing patterns motivate the use of autoregressive models, which are well-suited to capturing the temporal and logical progression of user interactions.
.

\textbf{Challenges. }While promising, these two-stage approaches face two significant challenges in real-time industrial recommendation systems. (1) \textit{The separation of the generator and evaluator severely hinders end-to-end training.} After the generator produces potential sequences, conventional two-stage approaches rely on an additional evaluator to select those that best align with user preferences. However, the separation impedes end-to-end training, resulting in a sharp increase in system complexity, misaligned optimization objectives, and limited generalization. (2) \textit{Autoregressive generators suffer from inference efficiency.} Autoregressive models adopt a sequential approach to generate target sequences item by item, resulting in slow inference as the time complexity increases linearly with the sequence length.

\textbf{Our Work.} To address the challenges mentioned above, we propose a Unified \textbf{G}enerative\footnote{In this paper, we primarily use the term "generative" to refer to autoregressive GPT-style models \cite{yang2024qwen2,abdin2024phi, openai2023gpt}, which are characterized by causal attention, next-token prediction. \label{footnote:generative}} Efficient \textbf{Re}ranking \textbf{F}ramework, named GReF. Specifically, we propose Gen-Reranker, an autoregressive model with a bidirectional encoder and a decoder that generates causal recommendation sequences through dynamic matching. Then we pre-train Gen-Reranker on the large-scale unlabeled item exposure order\footnote{The term "item exposure order" refers to the item sequence ultimately displayed to the user by the recommendation system. \label{footnote:item exposure order}} from the recommendation system to efficiently initialize the model parameters. We then develop specialized strategies to address  key challenges of the separation of the generator and evaluator and autoregressive inference efficiency:
\textbf{First, to unify sequence-level evaluation during training for end-to-end optimization }(\textbf{challenge 1}), we propose post-training Gen-Reranker through \textit{Rerank-DPO}, which constructs pair-wise objectives to compare user-preferred recommendation sequences with less-preferred ones, thereby directly integrating sequence-level user preferences during training without the need for an additional evaluator. 
\textbf{Second, to enable efficient autoregressive inference in real-time industrial scenarios }(\textbf{challenge 2}), we design \textit{Ordered Multi-token Prediction (OMTP)}, which trains Gen-Reranker to simultaneously generate multiple future items while preserving their order, thereby substantially reducing the latency of autoregressive inference in real-world recommendation systems.

To summarize, our contributions are listed as follows: 
\begin{itemize}
    \item We propose an end-to-end generative reranking framework that integrates the Gen-Reranker model into a unified training pipeline. This design effectively unifies traditional two-stage methods by bridging the gap between the generator and the evaluator, enabling end-to-end optimization.

    \item  We introduce OMTP, a novel method that simultaneously generates multiple future items while preserving their order. OMTP substantially improves the efficiency of autoregressive generation and facilitates seamless deployment of GReF in real-time recommendation systems.

    \item We conduct extensive offline experiments on both a public dataset and a real-world industrial dataset from Kuaishou, demonstrating that GReF outperforms state-of-the-art reranking methods while maintaining latency nearly comparable to non-autoregressive models. \textbf{Notably, GReF has been deployed at Kuaishou, delivering significant improvements across multiple metrics.}
\end{itemize}

\section{RELATED WORK}
\subsection{Reranking in Recommendation Systems}
Reranking in recommendation aims to reorder an input item list by modeling inter-item correlations to maximize user feedback. Reranking models typically take the entire list (listwise context) as input and output a reordered sequence. Existing methods can be broadly divided into one-stage and two-stage approaches.

One-stage methods \cite{pang2020setrank,pei2019personalized,xi2021context,ai2018learning,bello2018seq2slate} treat reranking as a retrieval task, where the top-$k$ items are greedily selected based on scores from a context-aware model. For example, PRM \cite{pei2019personalized} and DLCM \cite{ai2018learning} use Transformers \cite{vaswani2017attention} or RNNs \cite{chung2014empirical} to model item dependencies and predict click probabilities. However, modifying the item order alters inter-item interactions, making the original scores less reliable.

Two-stage methods \cite{feng2021grn,shi2023pier,lin2023discrete,xi2021context,feng2021revisit,ren2024non} adopt a generator-evaluator framework: the generator produces candidate sequences, and the evaluator selects the best one. Generators are typically based on heuristics (e.g., beam search or item swapping) \cite{xi2021context,shi2023pier,feng2021revisit}, or generative models \cite{bello2018seq2slate,feng2021grn,gong2022real,zhuang2018globally,lin2023discrete,ren2024non}. PRS \cite{feng2021revisit} uses beam search to generate candidates and scores them with a permutation-wise model. PIER \cite{shi2023pier} filters top-$K$ candidates using permutation-level interest and employs attention modules for listwise CTR prediction. DCDR \cite{lin2023discrete} introduces a discrete diffusion model using step-wise operations (e.g., swaps), conditioned on expected user responses. NAR4Rec \cite{ren2024non} adopts a non-autoregressive model \cite{gu2017non} with unlikelihood training and contrastive decoding to enhance intra-list correlation. Building on this, NLGR incorporates neighbor lists to improve training. Despite their effectiveness, these two-stage methods face challenges in end-to-end optimization due to the separation between generation and evaluation.

\subsection{Autoregressive Modeling}
Autoregressive language models~\cite{radford2018improving,radford2021learning,brown2020language,openai2023gpt,achiam2023gpt,touvron2023llama,touvron2023llama2,dubey2024llama,team2023gemini,chowdhery2023palm,anil2023palm,bai2023qwen,lin2021graph,lin2019data} have laid a promising foundation for general-purpose AI. At their core lies the autoregressive generation paradigm—predicting the next token based on previous context—a simple yet powerful approach that scales effectively~\cite{kaplan2020scaling} and demonstrates strong generalization and memorization abilities~\cite{achiam2023gpt}.

Autoregressive generation is also gaining popularity in reranking tasks~\cite{bello2018seq2slate,yan2024llm4pr,gao2025llm4rerank,zhang2025enhancing}. Seq2Slate~\cite{bello2018seq2slate} employs a seq2seq model~\cite{sutskever2014sequence} to select items step by step. More recently, LLM4PR~\cite{yan2024llm4pr} and LLM4Rerank \cite{gao2025llm4rerank} leverage zero-shot LLMs with Chain-of-Thought prompting for reranking. While effective, their high inference cost makes them impractical for real-time recommendation.

To address these issues, we propose a novel autoregressive reranking generator equipped with an efficient training framework. Our method effectively models item-level causality and user intent while remaining suitable for real-time recommendation scenarios.

\section{PRELIMINARY} 
\subsection{Reranking Task}

In multi-stage real-time recommendation systems, reranking plays a crucial role in generating the ultimate list of recommended items \cite{liu2022neural}. Given a set of candidates $X = \{ x_1, x_2, \dots, x_m \}$ from the preceding stage, the goal of reranking is to produce an ordered sequence of recommendations $Y = \{ y_1, y_2, \dots, y_n \}$ that maximizes the likelihood of favorable feedback from user $u$. Here, $m$ denotes the number of candidates, and $n$ represents the length of the recommended sequence. Typically, $m$ is much larger than $n$, with $m$ ranging from several tens to hundreds, while $n$ is usually less than 10.

\subsection{Autoregressive Reranking}
\label{pre:ar}
Given a set of candidate items denoted as $X = \{ x_1, x_2, \dots, x_m \}$, autoregressive reranking models factorize the joint probabilities over possible generated sequences $Y = \{ y_0, y_1, \dots, y_{n+1} \}$ into the product of conditional probabilities:
\begin{equation}
    p{(Y|X; \theta)} = \prod_{t=1}^{n+1} p(y_t|y_{0:t-1},x_{1:m};\theta),
\label{equ_ar}
\end{equation}
where $n$ represents the length of the recommended sequence. The special tokens $y_0$ (e.g., {\tt [BOS]}) and $y_{n+1}$ (e.g., {\tt [EOS]}) signify the beginning and end of the target sequence, respectively. In contrast to natural language processing (NLP) tasks, where the sequence length may vary, the sequence length in autoregressive reranking is fixed.

In training, the objective is to optimize the following likelihood using a cross-entropy loss at each timestep:
\begin{equation}
   \mathcal{L}_{\text{AR}} = -\log p_{\text{AR}}(Y|X; \theta) = -\sum_{t=1}^{n+1}\log p(y_t|y_{0:t-1},x_{1:m};\theta).
\end{equation}

In inference, autoregressive reranking models generate the target sequence sequentially through next-item prediction, effectively modeling the distribution of the target sequence. This property makes them particularly well-suited for reranking tasks, especially given the large combinatorial space of possible permutations. 

\subsection{Direct Preference Optimization (DPO)} \label{pre:direct}
DPO \cite{Rafailov2023DirectPO} is one of the most popular preference alignment methods in the LLM post-training stage. Rather than explicitly learning a reward model \cite{Ouyang2022TrainingLM}, DPO reparameterizes the reward function $r$ by leveraging a closed-form expression  with the optimal policy:

\begin{equation}
r(x,y) = \beta \log \frac{\pi_\theta(y \mid x)}{\pi_{\text{ref}}(y \mid x)} + \beta \log Z(x),
\end{equation} where $\pi_\theta$ is the policy model, $\pi_{\text{ref}}$ is the reference policy, typically the supervised fine-tuned (SFT) model, and $Z(x)$ is the partition function.  By incorporating this reward formulation into the Bradley-Terry (BT) ranking objective~\cite{Bradley1952RankAO}, $p(y_w \succ y_l \mid x) = \sigma \left( r(x, y_w) - r(x, y_l) \right)$, DPO represents the probability of preference data with the policy model, yielding the following objective:

\begin{equation} \label{pre:dpo}
     -\min_{\pi_\theta} \mathbb{E}_{(x, y_w, y_l) \sim \mathcal{D}}\left[ \log \sigma \left( \beta \log \frac{\pi_\theta(y_w \mid x)}{\pi_{\text{ref}}(y_w \mid x)} - \beta \log \frac{\pi_\theta(y_l \mid x)}{\pi_{\text{ref}}(y_l \mid x)}\right) \right],
\end{equation}
where $\pi_\theta$ is the policy model, $\pi_{\text{ref}}$ is the reference policy, and $(x, y_w, y_l)$ represents preference pairs consisting of the prompt $x$, the winning response $y_w$, and the losing response $y_l$ from the preference dataset $\mathcal{D}$.

\section{GENERATIVE RERANKING FRAMEWORK}
\begin{figure*}[h]
  \centering

  \includegraphics[width=\textwidth]{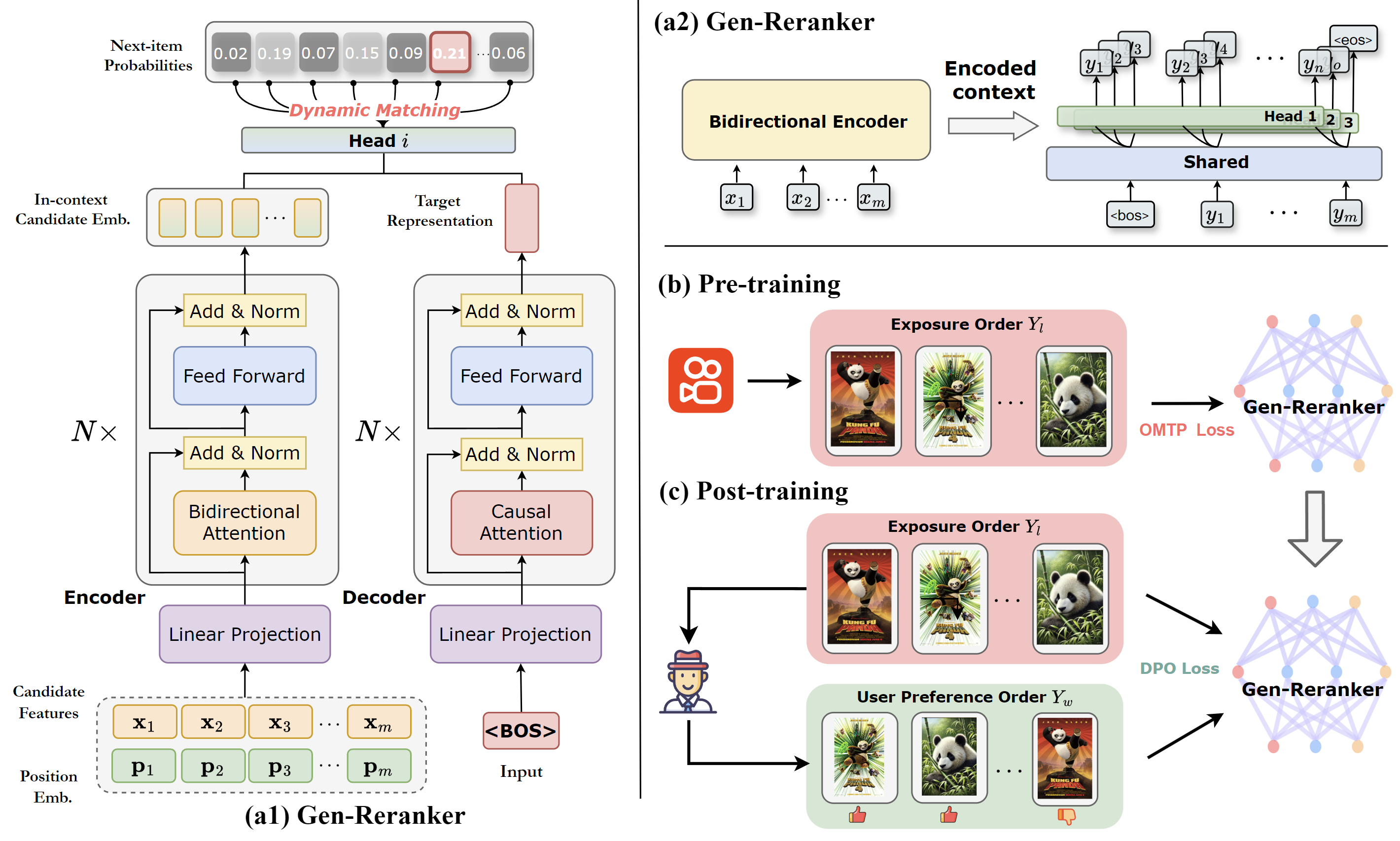}
  \caption{An overview of GReF. GReF consists of Gen-Reranker, which integrates a bidirectional encoder and a dynamic autoregressive decoder with dynamic matching (a1). Specifically, (a2) provides a detailed explanation of how Gen-Reranker generates multiple future items  simultaneously through OMTP. Additionally, GReF employs an effective training framework that combines pre-training (b) and post-training (c).} 
  \label{fig:framework}
  \vspace{0em}
\end{figure*}

In this section, we provide a detailed introduction to our unified generative framework for effective reranking illustrated in Figure \ref{fig:framework}. We first describe the structure of our Gen-Reranker model in Section \ref{sec_method:AR}, which generates the causal reranking sequence in an autoregressive manner. We then explore the training mechanism of Gen-Reranker. As detailed in Section \ref{sec_method:pre-train}, we first pre-train Gen-Reranker to achieve high-quality parameter initialization and enhance generalization. Building on this, Section \ref{sec_method:post-train} introduces \textit{Rerank-DPO}, which further aligns the model with user preferences and improves personalization. Finally, to enable efficient autoregressive generation, we introduce Ordered Multi-Token Prediction (OMTP), which generates items simultaneously in a structured order in Section \ref{sec_method:omtp}.

\subsection{Gen-Reranker}  \label{sec_method:AR}


Autoregressive reranking generates recommendations item by item, capturing complex causal dependencies for more personalized and accurate recommendations. While promising, the effectiveness of autoregressive reranking is hindered by the billion-scale and evolving item vocabularies \footnote{In autoregressive reranking, an item is equivalent to a token in NLP.} in recommendation systems, in contrast to the relatively static 100K-scale vocabularies in language models \cite{brown2020language}. To address this, we propose \textit{Gen-Reranker}, which comprises two key components: an encoder that leverages bidirectional attention to extract embeddings for each candidate in current session, and a decoder that autoregressively generates the next-item representation and then dynamically matches it with the candidate embeddings to predict the next item, thereby eliminating the need to consider billions of candidates across the entire recommendation system.

\textbf{Candidate Items Embeddings.} We employ a bidirectional transformer \cite{vaswani2017attention} 
 as encoder to generate target-aware embeddings for the candidate items.





Considering a candidate item set $X = \{ x_1, x_2, \dots, x_m \}$, each item $x_i$ is represented by item feature $\mathbf{x}_i \in \mathbb{R}^d$. These features are stacked into a matrix $\mathbf{X} \in \mathbb{R}^{m \times d}$. Additionally, a position embedding $\mathbf{p}_i \in \mathbb{R}^d$ is randomly initialized for each position $i$, forming a matrix $\mathbf{P} \in \mathbb{R}^{m \times d}$. Note that this position order comes from the previous ranking stage. The input representations are then obtained by summing the item features $\mathbf{X}$ and their corresponding position embeddings $\mathbf{P}$. Subsequently, we input them into the bidirectional encoder. Finally, the final hidden state  $\mathbf{Z} = \{\mathbf{z}_1, \mathbf{z}_2, \dots, \mathbf{z}_m\}$ serves as the candidate embeddings, seamlessly integrating in-context item information.

\textbf{Dynamic Multi-Item Prediction.} To achieve high efficiency in real-time autoregressive generation, we propose \textit{dynamic multi-item prediction}. We replace the standard output layer over the full vocabulary with a concatenation of embeddings corresponding to the candidate items to be reranked. This allows the model to generate outputs solely from the in-context candidate set, eliminating the need to compute over the entire billion-scale item pool in recommendation systems. Moreover, instead of generating one item at a time, we adopt multi-item prediction to enable parallel decoding of multiple future items, significantly improving inference efficiency.

To be more specific, we replace the output layer’s weights with candidate embeddings $\mathbf{Z}$ from encoder to create dynamic in-context item vocabularies.  During decoding, we compute similarity scores between the decoder's last hidden state $\mathbf{h}$ and the linear layer weight $\mathbf{Z}$ to obtain logits and then apply the softmax function to compute the next-item probabilities. The process is formulated as follows: 
\begin{equation} 
p_\theta(y_t|y_{0:t-1}) =\frac{\exp \left(\mathbf{h}_{t}^{\top} \mathbf{z}_{y_t}\right)}{\sum_{i=0}^{m} \exp \left(\mathbf{h}_{t}^{\top} \mathbf{z}_{i}\right)} ,
\label{eq:p_theta}
\end{equation}
where $\mathbf{h}_{t}$ and $\mathbf{z}_{y_t}$ represent the final hidden state and item embedding for target item $y_t$, respectively.

\subsection{Pre-training on Recommender World Knowledge} \label{sec_method:pre-train}
To train reranking models, traditional approaches \cite{pei2019personalized, feng2021revisit, shi2023pier, lin2023discrete} typically rely on data with user feedback, such as clicks or likes. While promising, user feedback in real-world recommendation systems is often extremely sparse \cite{liu2020long}. Training solely on these samples can lead to overfitting and hinder generalization.


To address this, similar to how LLMs are pre-trained on vast unlabeled data from the internet, we propose pre-training the model on the item exposure order from the recommendation system. This approach is motivated by two key reasons. First, training data is sufficient. Using item exposure order allows the training samples to include all item data, rather than just the items that have been interacted with by users. Second, learning from the item exposure order provides the model with high-quality parameter initialization. A well-trained multi-stage recommendation system, such as YouTube or Kuaishou, incorporates broader world knowledge beyond just user interactions, including expert rules and contextual information. This means that pre-training on the item exposure order generated from modern recommendation systems not only allows the AR model to accurately and comprehensively capture users' interests but also enhances its generalization and robustness.

Specifically, given multiple candidate item sets $\mathcal{X}_{train} = \{X_i|X_i = \{ x_1, x_2, \dots, x_m \}\}_{i=1}^K$, We can obtain the ordered exposure sequences $\mathcal{Y}_{train} = \{Y_i|Y_i = \{ y_1, y_2, \dots, y_n \}\}_{i=1}^K$ presented to the user from the recommendation system,  where $\mathcal{Y}_{train}$ is a subset of $\mathcal{X}_{train}$. Subsequently, we add a special \texttt{[BOS]} token at the beginning and a \texttt{[EOS]} token at the end of each $Y_i$ to obtain target sequences $\mathcal{Y}_{train} = \{Y_i \mid Y_i = \{y_0, y_1, \dots, y_n, y_{n+1}\} \}_{i=1}^K$. Finally, we pre-train our autoregressive reranking model on $\mathcal{Y}_{train}$ through cross-entropy loss as described in Section \ref{pre:ar}: 
\begin{equation} \label{pre-train}
   \mathcal{L}_{pre-train} = -\frac{1}{K}\sum_{\mathcal{Y}_{train}}\sum_{t=1}^{n+1}\log p_\theta(y_t|y_0,y_1,\cdots,y_{t-1}),
\end{equation}
where $p_\theta$ denotes an item distribution predictor with a
model parameterized by $\theta$.





\subsection{Post-training on User Preferences} \label{sec_method:post-train}

%


A key remaining challenge is how to further integrate user preferences into the model. Conventional two-stage approaches address this by using the evaluator to select the sequences generated from the generator that best align with user preferences. However, this pattern hinders end-to-end training, leading to significantly increased system complexity, misaligned optimization objectives, and limited generalization.


To address this, inspired by DPO \cite{rafailov2024direct} in LLMs (with more details provided in Section \ref{pre:direct}), we propose \textit{Rerank-DPO}, which first constructs preference pairs based on item exposure order and user feedback, and then forms a pairwise optimization objective from these pairs, directly integrating sequence-level user preferences into the model.

Specifically, given an exposure order $Y = \{y_1, y_2, \dots, y_{n}\}$, we first assess an item's personalization score based on two criteria: (1) original exposure position and (2) user feedback. Utilizing these, our objective is to construct winning and losing sequences that capture both user and recommendation system preferences, ensuring that the integration of user preferences does not interfere with the knowledge acquired during the pre-training stage.  Thus we can define the personalization score of item $y_i$ as: 
\begin{equation}
   S_i = \alpha * \frac{1}{P_i} + \gamma * U_i,
\end{equation}
where $P_i$ represents the position of $y_i$ in the exposure order, and $U_i$ denotes user feedback on $y_i$, such as clicks, with values of 0 or 1. $\alpha$ and $\gamma$ denote the balance hyper-parameter. Afterward, we can rank items in the exposure order $Y$ according to the score to get the personalized sequence $Y_{w}$. For example, if the user clicks on item $y_2$ and $y_5$ in $Y$, we can derive a personalized sequence  $Y_w = \{y_2, y_5, y_1, y_3, y_4, \dots, y_n\}$. Finally, if $Y_w$ differs from the original exposure order, we treat $Y_w$ as the winning sequence and exposure order, denoted as $Y_l$, as the losing sequence, respectively. 

Based on this, we can construct the preference dataset $\mathcal{Y}_{post-train} = \{(Y_w, Y_l)_i\}_{i=1}^K$ and post-train our pre-trained Gen-Reranker through DPO to align with user preferences. The loss function is defined as follows: 
\begin{align}
\label{pre:dpo}
&\mathcal{L}_{dpo} =  \\ \notag
     &-\min_{\pi_\theta} \mathbb{E}_{(Y_w, Y_l) \sim \mathcal{Y}_{post-train}}\left[ \log \sigma \left( \beta \log \frac{\pi_\theta(Y_w)}{\pi_{\text{ref}}(Y_w)} - \beta \log \frac{\pi_\theta(Y_l)}{\pi_{\text{ref}}(Y_l)}\right) \right],
\end{align} 
where $\pi_\theta$ is the pre-trained Gen-Reranker, $\pi_{\text{ref}}$ is the frozen version, and $\pi_\theta(Y_w)$, $\pi_{\text{ref}}(Y_w)$, $\pi_\theta(Y_l)$, and $\pi_{\text{ref}}(Y_l)$ represent the summed log probabilities of sequences $Y_w$ and $Y_l$ on $\pi_\theta$ and $\pi_{\text{ref}}$, respectively. $\beta$ is the hyper-parameter.

\subsection{Ordered Multi-token Prediction} \label{sec_method:omtp}
To achieve high inference efficiency in real-time autoregressive generation for real-world applications, we propose Ordered Multi-token Prediction (OMTP), which trains Gen-Reranker with $n$ output heads to generate multiple future items in a single forward pass while preserving their order.

Specifically, given the observed context $y_{0:t-1}$, we first encode it into the latent representation $h_{0:t-1}$ via the shared trunk Gen-Reranker, which is then fed into the $n$ output heads to generate predictions for the future items. This design leads to the following cross-entropy loss \cite{gloeckle2024better}:

\begin{align} \label{eq:loss-n}
\mathcal{L}_n
&= - \sum_t \log p_\theta(y_{t:t+n-1} \mid h_{0:t-1}) \cdot p_\theta(h_{0:t-1} \mid y_{0:t-1}) \notag \\
&= - \sum_t \sum_{i=0}^{n-1} \log p_\theta(y_{t+i} \mid h_{0:t-1}) \cdot p_\theta(h_{0:t-1} \mid y_{0:t-1}),
\end{align}
where $p_\theta(y_{t+i} \mid y_{0:t-1})$ from head $i$ is computed through Eq. \ref{eq:p_theta}. 


To further ensure that the multiple items generated by different heads follow a sequential order, which is crucial for modeling user behavior in short-video recommendation, we construct positive and negative item pairs and apply a pairwise loss to capture their relative order. Specifically, at step $t$, we enumerate permutations of the outputs from $n$ heads. Suppose a permutation $Y_+$ receives a higher score, measured by a scoring function $S$ such as NDCG based on user clicks. In that case, we treat it as a positive example and another as negative $Y_-$, thereby guiding the model to learn the correct item sequence:
\begin{equation} \label{loss_o}
    \mathcal{L}_o = - \sum_{t, S(Y_+)>S(Y_-)} \log \sigma(P_{\theta}(Y_+ \mid y_{0:t-1})-P_{\theta}(Y_-\mid y_{0:t-1})).
\end{equation}

The OMTP loss is defined as:
\begin{align}
    \mathcal{L}_{omtp} = \lambda_1*\mathcal{L}_n + \lambda_2*\mathcal{L}_o, 
\end{align}
where $\lambda_1$ and $\lambda_2$ are hyper-parameters. 

\subsection{Optimization and Inference}
In the pre-training stage, we train Gen-Reranker on large-scale unlabeled item exposure sequences optimized via the loss function $\mathcal{L}{\text{omtp}}$.
Subsequently, we post-train Gen-Reranker using Rerank-DPO, supervised by the loss $\mathcal{L}{dpo}$. During inference, OMTP predicts multiple items in parallel. To avoid duplicates, we apply a binary mask at each step to exclude previously selected items, enabling efficient autoregressive decoding for real-time industrial applications.

\section{EXPERIMENTS}
In this section, we present extensive offline experiments and online A/B tests to validate the effectiveness of GReF. We first introduce the baselines and implementation details in Section \ref{sec:settings}. In Section \ref{sec:offline}, we compare GReF with existing baselines in terms of performance and inference time. We also conduct ablation studies to assess the contributions of different training stages and OMTP. To further demonstrate the effectiveness of GReF in real-time recommendation systems, we perform online A/B tests to evaluate our proposed methods in Section \ref{sec:online}.

\subsection{Experiment settings}
\label{sec:settings}
\subsubsection{Datasets} To validate the effectiveness of our framework, we conduct extensive experiments on both public and industrial datasets:

\begin{itemize}
\item \textbf{Avito}\footnote{https://www.kaggle.com/c/avito-context-ad-clicks/data}: A public dataset of user search logs from avito.ru, containing over 53 million lists, 1.3 million users, and 36 million ads. Each sample is a search page with multiple ads. Logs from the first 21 days are used for training and the last 7 days for testing. Each sequence has 5 ads. The task is to predict item-wise click-through rates given list-wise inputs.

\item \textbf{Kuaishou}: Collected from the Kuaishou short-video app with over 300 million daily active users. Each sample is a user request containing user features, 30 candidate items, and 10 exposed items. The dataset includes 300 million users, 733 million items, and 252 million requests. The task for the Kuaishou dataset is to predict whether an item is chosen to be one of the exposed 10 items.
\end{itemize}

\subsubsection{Baselines} We compare our model with 7 state-of-the-art reranking methods. We select DNN and DCN as pointwise baselines, PRM as a one-stage listwise baseline, and Edge-Rerank, PIER, Seq2Slate, and NAR4Rec as two-stage baselines. A brief overview of these methods is provided below:

\begin{itemize}
    \item \textbf{DNN} \cite{covington2016deep}: DNN a multi-layer perception for click-through rate prediction.
    \item \textbf{DCN} \cite{wang2017deep}:  DCN enhances DNNs with feature crossing at each layer.
    \item \textbf{PRM} \cite{pei2019personalized}: PRM uses self-attention to model item correlations and ranks items by predicted scores.
    \item \textbf{Edge-Rerank} \cite{gong2022real}: Edge-Rerank generates a context-aware sequence using adaptive beam search applied to the estimated scores.
    \item \textbf{Seq2Slate} \cite{bello2018seq2slate}:  Seq2Slate leverages pointer networks to select the next item based on previously chosen ones.
    \item \textbf{PIER} \cite{shi2023pier}: PIER filters top-k candidates via hashing, then jointly trains a generator and evaluator for better permutations.
    \item \textbf{NAR4Rec} \cite{ren2024non}: NAR4Rec combines a non-autoregressive generator with unlikelihood training, contrastive decoding, and a reranking evaluator.
\end{itemize}

\subsubsection{Implementation Details} We implement our model in Tensorflow \cite{abadi2016tensorflow} and perform all experiments on four NVIDIA T4 GPU. We configure the Bidirectional Encoder and Dynamic Autoregressive Decoder in our Gen-Reranker with $4$ stacked Transformer layers each. For Rerank-DPO, we set the balance hyperparameters to $\alpha = 1$ and $\gamma = 1$, treating clicks as user feedback with values of 0 or 1 while adopting $\beta = 0.1$ to control the scaling of the reward difference. For OMTP, we set $n=4$ to match the Kuaishou app's UI, which displays four videos per screen. Both $\alpha$ and $\beta$ are set to 1. During training, we use Adam \cite{kingma2014adam} with an initial learning rate $\eta = 5e^{-6}$ to update the parameters, and the batch size is set to $1,024$. We first pre-train Gen-Reranker and then post-train it using Rerank-DPO. We run our model and other baseline methods five times with different random seeds. Additionally, we either optimize the parameters or follow the settings in the original papers to achieve the best performance for the baseline methods.

\subsubsection{Metrics} Following previous work \cite{shi2023pier, lin2023discrete, ren2024non}, we evaluate different methods in offline experiments using AUC and NDCG on both the Avito and Kuaishou datasets.



\subsection{Offline experiments}
\label{sec:offline}
\subsubsection{Performance comparison}  Here we show the results of our proposed method GReF. As shown in Table \ref{tab:combined_results_updated},  GReF outperforms all baseline methods on both the Avito and KuaiShou datasets across key evaluation metrics. On the Avito dataset, GReF achieves the highest AUC, about 1.5\% higher than the next best model, and the best NDCG, approximately 0.8\% higher than PIER. Similarly, on the KuaiShou dataset, GReF leads with an AUC, about 1.4\% higher than the best-performing baseline, and an NDCG, around 0.5\% better than other methods. The consistent performance across both datasets highlights GReF's robustness and ability to enhance recommendation quality, making it a highly effective approach in practical scenarios.


\begin{table}[ht]
\centering
\caption{Comparison between GReF and baseline methods on the Avito and KuaiShou datasets. The best results on each dataset are highlighted in bold.}
\label{tab:combined_results_updated}
\begin{tabular}{l|cc|cc}
\toprule
\multirow{2}{*}{\textbf{Method}} 
& \multicolumn{2}{c|}{\textbf{Avito}} 
& \multicolumn{2}{c}{\textbf{KuaiShou}} \\ 
\cmidrule(lr){2-3} \cmidrule(lr){4-5}
& AUC & NDCG & AUC & NDCG \\ 
\midrule
DNN         & 0.6614          & 0.6920          & 0.6866          & 0.7122 \\
DCN         & 0.6623          & 0.7004          & 0.6879          & 0.7116 \\
PRM         & 0.6881          & 0.7380          & 0.7119          & 0.7396 \\
Edge-rerank & 0.6953          & 0.7203          & 0.7143          & 0.7401 \\
PIER        & 0.7109          & 0.7401          & 0.7191          & 0.7387 \\
Seq2Slate   & 0.7034          & 0.7225          & 0.7165          & 0.7383 \\ 
NAR4Rec     & 0.7234          & 0.7409          & 0.7254          & 0.7425 \\ \midrule
GReF        & \textbf{0.7384} & \textbf{0.7478} & \textbf{0.7387} & \textbf{0.7498} \\ 
\bottomrule
\end{tabular}
\end{table}



\subsubsection{Inference Time comparison}
We evaluate GReF’s efficiency by comparing inference times on the Kuaishou dataset (Table \ref{tab:inference_time}) using a Tesla T4 16G GPU, measuring average single-sample latency. Classical models like DNN and DCN are highly efficient due to shallow architectures, while PRM, ERK, and PIER incur moderately higher latency. The non-autoregressive NAR4Rec achieves 12.67 ms. GReF achieves a strong balance between accuracy and efficiency, outperforming all baselines in AUC and NDCG with an inference time of 12.97 ms, close to NAR4Rec and over 4× faster than Seq2Slate due to ordered multi-token prediction (OMTP). Without OMTP, GReF’s step-by-step decoding results in the highest latency. OMTP predicts multiple tokens per step, nearly halving forward passes and latency.

This highlights OMTP’s effectiveness in accelerating autoregressive inference, enabling GReF to combine generative expressiveness with real-time deployability.




\begin{table}[]
\caption{Inference time comparison between GReF and baselines on the Kuaishou dataset. Note that ERK denotes Edge-Rerank.}
\begin{adjustbox}{width=\linewidth}
\begin{tabular}{@{}l|ccccccccc@{}}

\toprule
               & \textbf{DNN}           & \textbf{DCN}  & \textbf{PRM}   & \textbf{ERK} & \textbf{PIER}  & \textbf{NAR4Rec} &\textbf{Seq2Slate} & \textbf{GReF (w/o OMTP)}  & \textbf{GReF}  \\ \midrule
\textbf{Inf. Time (ms)} & 8.03 & 8.25 & 10.96 & 10.45       & 13.69 & 12.67 & 67.34  & 24.29 & \textbf{12.97} \\ \bottomrule

\end{tabular}
\label{tab:inference_time}
\end{adjustbox}
\vspace{-5mm}
\end{table}

\subsubsection{Ablation Study of Training Stage} To investigate the contributions of different training stages, we conduct the ablation studies and present AUC and NDCG results in Table \ref{tab:combined_ablation}. We can find that (1) Pre-training on exposure order provides the model with high-quality parameter initialization, effectively capturing user interest in many cases. (2) Post-training is essential for further integrating user preferences, leading to an improvement of 0.63\% in AUC and 0.45\% in NDCG. (3) Relying solely on post-training leads to a significant degradation in model performance. One possible explanation is that directly applying reinforcement learning-based DPO on user feedback data in a cold-start scenario can easily result in training instability and even collapse.


\begin{table}[ht]
\footnotesize
\centering
\caption{Ablation studies of training stages and OMTP loss in the pre-training stage on the Kuaishou dataset.}
\label{tab:combined_ablation}
\begin{tabular}{@{}cc|cc||cc|cc@{}}
\toprule
\multicolumn{4}{c||}{\textbf{Training Stages}} & \multicolumn{4}{c}{\textbf{OMTP Loss}} \\ \midrule
Pre-training & Post-training & AUC    & NDCG   & $\mathcal{L}_n$ & $\mathcal{L}_o$ & AUC    & NDCG   \\ \midrule
$\checkmark$   &                & 0.7361          &  0.7474          &                &                & 0.7324          & 0.7453          \\
               & $\checkmark$   & 0.6832          & 0.7103          & $\checkmark$   &                & 0.7373          & 0.7484          \\ \midrule
$\checkmark$   & $\checkmark$   & \textbf{0.7387} & \textbf{0.7498} & $\checkmark$   & $\checkmark$   & \textbf{0.7387} & \textbf{0.7498} \\ \bottomrule
\end{tabular}
\end{table}

\subsubsection{Ablation Study of OMTP} To further evaluate the effectiveness of the OMTP objectives $\mathcal{L}_n$ (Eq.\ref{eq:loss-n}) and $\mathcal{L}_o$ (Eq.\ref{loss_o}) during pre-training, we conduct ablation studies, with results presented in Table~\ref{tab:combined_ablation}. Note that using either loss $\mathcal{L}_n$  and $\mathcal{L}_o$ alone corresponds to single-head autoregressive pre-training as defined in Eq.~\ref{pre-train}, and all models undergo post-training. We highlight the following findings: (1) The model, which predicts multiple tokens in a single forward pass and is trained using $\mathcal{L}_n$ alone, achieves performance comparable to the full autoregressive framework. This demonstrates that the MTP paradigm can serve as a strong alternative to traditional autoregressive methods. (2) Adding the ordered loss $\mathcal{L}_o$ on top of $\mathcal{L}_n$ leads to further improvements, confirming that explicitly enforcing a preference-consistent generation order significantly enhances model performance. This indicates that OMTP not only greatly accelerates inference compared to autoregressive approaches but also maintains, or even improves, recommendation accuracy.

\subsection{Online experiments}
\label{sec:online}
To further demonstrate our effectiveness, we conduct online A/B experiments on the Kuaishou app with 8\% of the entire production traffic over one week. The online baseline is NAR4Rec \cite{ren2024non}.

\subsubsection{Experimental Results} The experiments have been launched on the system over one week, and the result is listed in Table~\ref{tab:online}. GReF outperforms NAR4Rec by a large margin. To be more specific, the Views increase by +0.33\%, indicating better overall visibility of recommended content. Long Views rise by +0.42\% with a p-value less than 0.01 and a Confidence Interval of [0.31\%, 0.52\%], reflecting enhanced user engagement and prolonged content interaction. Likes show a +1.19\% improvement, suggesting higher user satisfaction and content appreciation. Most notably, Forwarding (shares) increases by +2.98\%, highlighting the model's effectiveness in encouraging content sharing and expanding its reach. Additionally, Comments increase by +1.78\%, indicating that users are more inclined to engage in discussions and express their opinions on the recommended content. This suggests that GReF not only improves content visibility and engagement but also fosters a more interactive and participatory user experience.

\begin{table}[ht]
\caption{The experimental results from Online A/B testing. All values are the relative improvements of GReF. For the online A/B test in Kuaishou, the 0.2\% increase in views and long view, and 0.5\% in user interactions (like, forward, comment) are very significant improvements in our system.} 
\begin{tabular}{ccccc}
\toprule
Views & Long Views & Likes & Forwards & Comments \\
\midrule
    +0.33\% & +0.42\%  & +1.19\% & +2.98\% &  +1.78\%\\
\bottomrule
\end{tabular}
\label{tab:online}
\vspace{-3mm}
\end{table}

\section{Conclusion} 

In this paper, we provide an overview of the current formulation and challenges associated with reranking in recommendation systems. We emphasize the significance of autoregressive reranking and pioneer the deployment of the autoregressive reranking generator, GenReranker, in real-time recommendation systems. Building on this, we propose GReF, a novel generative reranking framework with end-to-end training. By seamlessly integrating the Gen-Reranker model into an optimized pipeline, GReF not only boosts the effectiveness of autoregressive reranking but also enables efficient real-time deployment. To validate its effectiveness, we conduct extensive offline and online A/B experiments, and the comparison with state-of-the-art reranking methods demonstrates the superiority of our approach.  In the future, our future work will focus on further enhance the capabilities of GReF.

\section{GenAI Usage Disclosure}

The authors affirm that no generative artificial intelligence (GenAI) tools were used in the preparation of this manuscript. All text, figures, tables, code, and other scholarly contributions were created solely by the authors without the assistance of generative AI software. The authors take full responsibility for the originality, accuracy, and integrity of the content.



\end{document}